\documentclass[sigconf]{acmart}
\pdfoutput=1

\copyrightyear{2022}
\acmYear{2022}
\setcopyright{acmcopyright}\acmConference[KDD '22]{Proceedings of the 28th ACM SIGKDD Conference on Knowledge Discovery and Data Mining}{August 14--18, 2022}{Washington, DC, USA}
\acmBooktitle{Proceedings of the 28th ACM SIGKDD Conference on Knowledge Discovery and Data Mining (KDD '22), August 14--18, 2022, Washington, DC, USA}
\acmPrice{15.00}
\acmDOI{10.1145/3534678.3539370}
\acmISBN{978-1-4503-9385-0/22/08}

\usepackage{multirow}
\usepackage{amsmath}
\usepackage{graphicx}
\usepackage{enumitem}
\usepackage{subfigure}
\usepackage{caption}
\usepackage[linesnumbered,ruled,vlined]{algorithm2e}
\SetKwInput{KwInput}{Input}                
\SetKwInput{KwOutput}{Output}              
\begin{document}

\title{CLARE: A Semi-supervised Community Detection Algorithm}

\newcommand{\sch}[1]{\textcolor{blue}{\bf \small [#1 --sch]}}
\newcommand{\zy}[1]{\textcolor{cyan}{\bf \small [#1 --zy]}}

\newcommand{\ours}{CLARE}
\newcommand{\first}{Locator}
\newcommand{\second}{Rewriter}

\author{Xixi Wu}
  \email{21210240043@m.fudan.edu.cn}
  \affiliation{%
    \institution{Shanghai Key Laboratory of Data Science, School of Computer Science, Fudan University}
    \country{China}
  }

\author{Yun Xiong}
\authornote{Corresponding author}
 \email{yunx@fudan.edu.cn}
  \affiliation{%
     \institution{Shanghai Key Laboratory of Data Science, School of Computer Science, Fudan University, Peng Cheng Laboratory, Shenzhen}
    \country{China}
  }
 
\author{Yao Zhang}
  \email{yaozhang@fudan.edu.cn}
  \affiliation{%
     \institution{Shanghai Key Laboratory of Data Science, School of Computer Science, Fudan University}
      \country{China}
  }
 
\author{Yizhu Jiao}
  \email{yizhuj2@illinois.edu}
  \affiliation{%
    \institution{University of Illinois at Urbana-Champaign}
    \country{United States}
  }
 
\author{Caihua Shan}
  \email{caihuashan@microsoft.com}
  \affiliation{%
    \institution{Microsoft Research Asia}
    \country{China}
  }
 
\author{Yiheng Sun}
  \email{elisun@tencent.com}
  \affiliation{%
    \institution{Tencent Weixin Group}
    \country{China}
  }
 
\author{Yangyong Zhu}
  \email{yyzhu@fudan.edu.cn}
  \affiliation{%
     \institution{Shanghai Key Laboratory of Data Science, School of Computer Science, Fudan University}
     \country{China}
  }
 
\author{Philip S. Yu}
  \email{psyu@uic.edu}
  \affiliation{%
    \institution{University of Illinois at Chicago}
    \country{United States}
}

\renewcommand{\shortauthors}{Xixi Wu et al.}

\begin{abstract}

Community detection refers to the task of discovering closely related subgraphs to understand the networks. However, traditional community detection algorithms fail to pinpoint a particular kind of community. This limits its applicability in real-world networks, e.g., distinguishing fraud groups from normal ones in transaction networks. Recently, semi-supervised community detection emerges as a solution. It aims to seek other similar communities in the network with few labeled communities as training data. Existing works can be regarded as seed-based: \textit{locate seed nodes and then develop communities around seeds}. However, these methods are quite sensitive to the quality of selected seeds since communities generated around a mis-detected seed may be irrelevant. Besides, they have individual issues, e.g., inflexibility and high computational overhead. To address these issues, we propose \textbf{CLARE}, which consists of two key components, \textbf{C}ommunity \textbf{L}ocator \textbf{a}nd Community \textbf{Re}writer. Our idea is that we can \textit{locate potential communities and then refine them}. Therefore, the community locator is proposed for quickly locating potential communities by seeking subgraphs that are similar to training ones in the network. To further adjust these located communities, we devise the community rewriter. Enhanced by deep reinforcement learning, it suggests intelligent decisions, such as adding or dropping nodes, to refine community structures flexibly. Extensive experiments verify both the effectiveness and efficiency of our work compared with prior state-of-the-art approaches on multiple real-world datasets.

\end{abstract}



\maketitle

\section{Introduction}

Networks are a powerful framework to represent rich relational information among data objects from social, natural, and academic domains\space \cite{Perozzi2014FocusedCA, Backstrom2006GroupFI}. A crucial step to understand a network is identifying and analyzing closely related subgraphs, i.e., communities. The research task of discovering such subgraphs from networks is known as the community detection problem\space \cite{Zhang2015IncorporatingIL}, which can reveal the structural patterns and inherent properties of networks.

\begin{figure}[t]
    \centering
    \includegraphics[width=7.5cm]{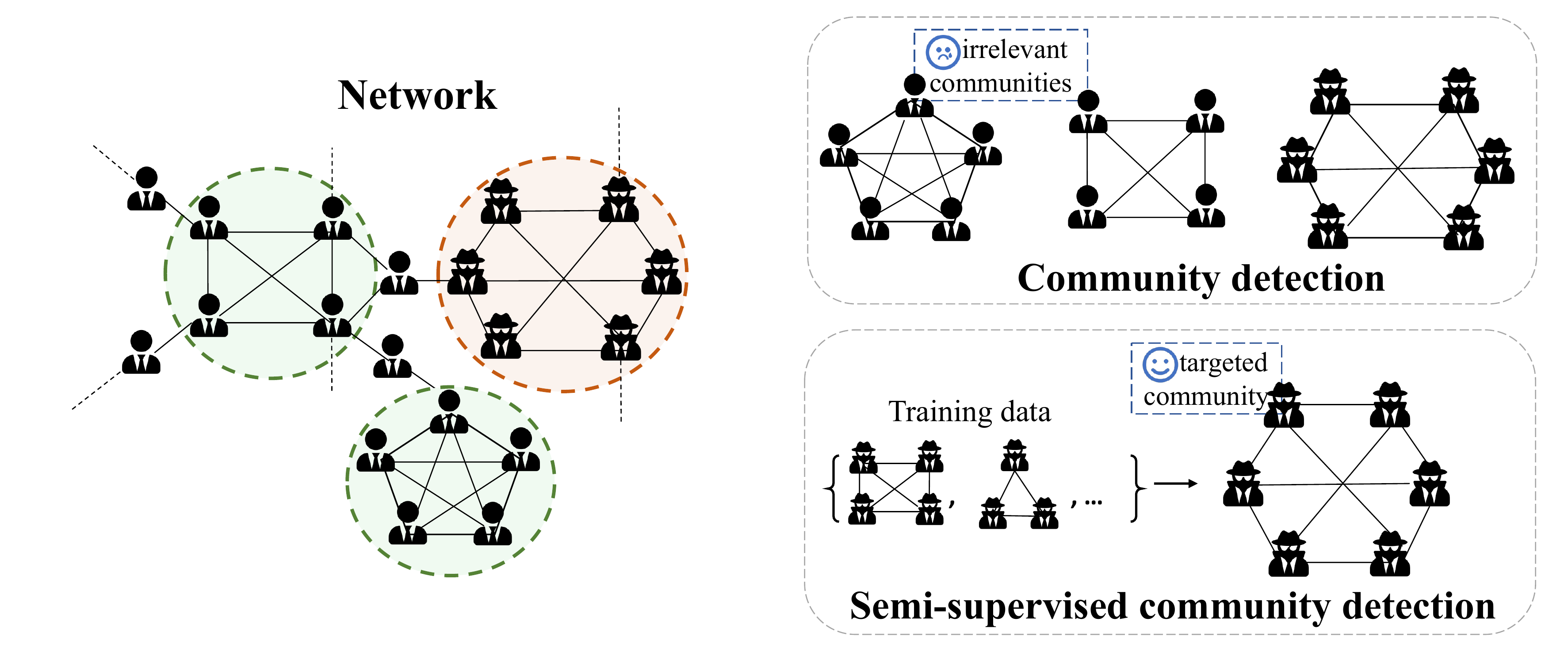}
    \caption{Tasks comparison. This is a subgraph of a trading network with two normal communities (green circles) and one fraud community (red circle). Traditional community detection methods may identify both kinds of communities. However, semi-supervised community detection methods can utilize some fraud groups to pinpoint the remaining fraud community.}
    \label{fig:task_comparison}
\end{figure}

However, traditional community detection algorithms are incapable of pinpointing a particular kind of community. In some scenarios, there are various types of communities in the same network, while people may only focus on a specific type, i.e., the targeted community. Traditional community detection methods cannot handle these situations, as they rely solely on overall structural information for inference  \cite{bigclam, Blondel_2008, Clauset_2004}, failing to capture the inherent features of some certain targeted communities. For example, as shown in Figure \ref{fig:task_comparison}, they cannot distinguish fraud groups from normal ones in transaction networks, and instead exhaustedly identify both kinds of communities.

Therefore, some researchers turn to semi-supervised settings to identify targeted communities. Formally, semi-supervised community detection is defined as utilizing certain communities as training data to recognize the other similar communities in the network. With such training data, it can implicitly learn the distinct features of the targeted community \cite{seal}. Semi-supervised community detection is a promising research problem with a wide range of real-world applications. For example, it can detect social spammer circles in social networks to avoid malicious attacks \cite{Hu2014OnlineSS}.

\begin{figure}[t]
    \centering
    \includegraphics[width=7cm]{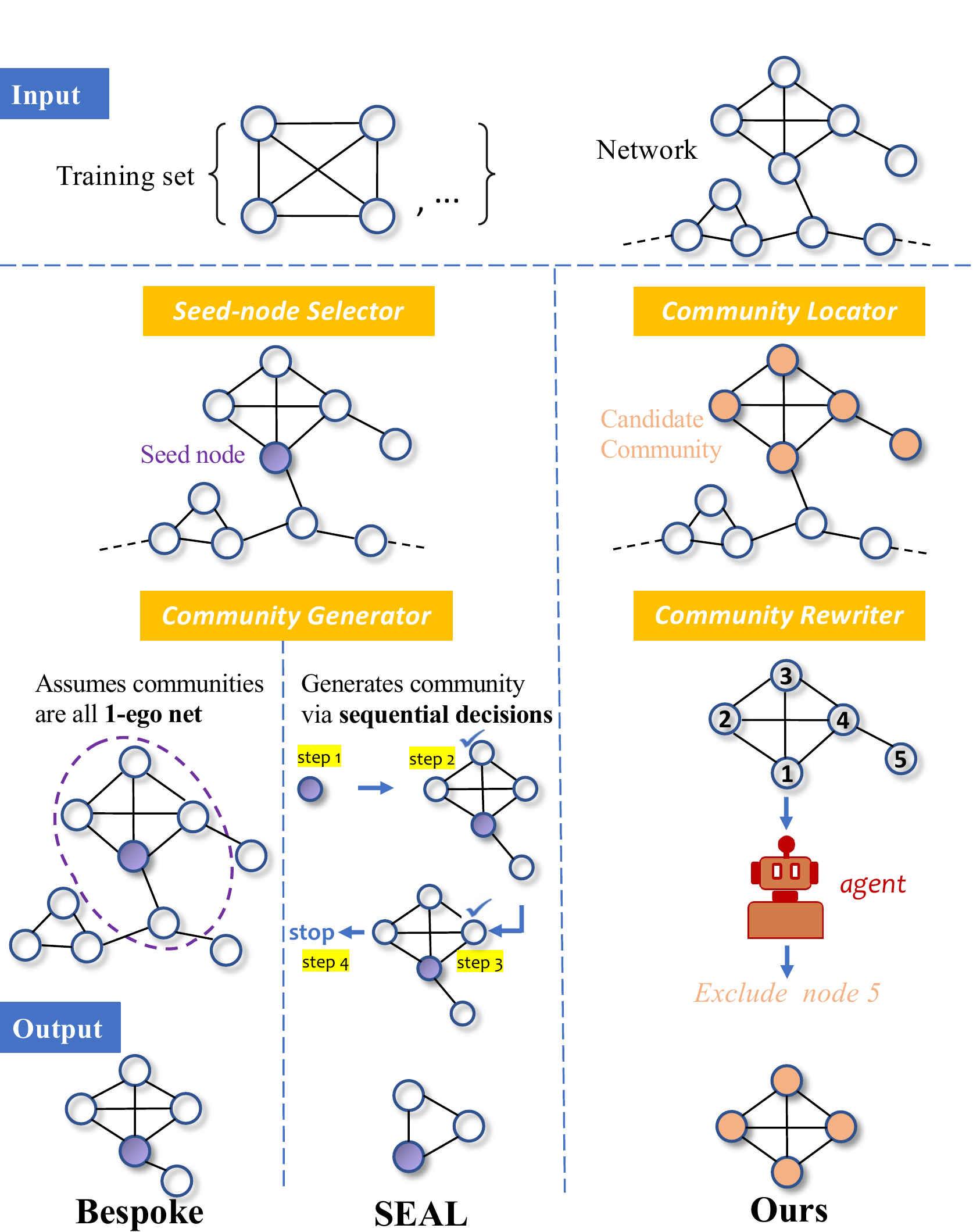}
    \caption{An illustration of Bespoke (left), SEAL (middle), and our proposed CLARE (right). Bespoke and SEAL first locate seed nodes and then develop communities around seeds. On the contrary, we propose a different solution: first locate potential communities and further rewrite them.}
    \label{fig:method_comparison}
\end{figure}

There are existing works on semi-supervised community detection, e.g., Bespoke \cite{bespoke} and SEAL \cite{seal}. As shown in Figure \ref{fig:method_comparison}, they can be generalized as seed-based methods: \textit{first locate prospective seed nodes (central nodes) in the network and then develop communities around seeds}. However, these methods are quite sensitive to the quality of selected seeds since communities generated around a mis-detected seed may be irrelevant. Besides, these methods have individual issues. For instance, Bespoke restricts community structures to 1-ego net, i.e., a central node along with its neighbors. This makes Bespoke inflexible as failing to generalize to communities with arbitrary structures. SEAL alleviates this issue by introducing generative adversarial networks (GAN). It formulates the generation of a community into sequential decisions process, as the generator of GAN would add one node in each step. However, it suffers from high computational overhead since developing a community in such way is quite expensive.

To solve the aforementioned challenges, we infer communities from a novel subgraph perspective: \textit{first locate potential communities and then refine them}. Specifically, we leverage the training communities to locate similar subgraphs via matching, and further adjust them based on their structural context. The benefits are threefold: (1) A subgraph carries more structural patterns as well as inherent features than a single seed node, promising to bring more precise positioning. (2) Predicting communities directly from a subgraph scale can avoid the expensive cost of generating communities from scratch. (3) With the introduction of global structural patterns, the refining process can further optimize located communities.

Inspired by the above insights, we propose a novel semi-supervised community detection framework, \textbf{CLARE}. It consists of two key components, \textbf{C}ommunity \textbf{L}ocator \textbf{a}nd Community \textbf{Re}writer. The community locator can quickly locate potential communities by seeking subgraphs that are similar to training ones. Specifically, we encode communities into vectors, measure the similarities between communities in the latent space, and then discover candidates based on the similarities with the nearest neighbors matching strategy. The community rewriter is proposed to further adjust those candidate communities by introducing global structural patterns. We frame such refinement process as a deep reinforcement learning task and optimize this process via policy gradient. For located communities, the rewriter provides two actions: adding outer nodes or dropping existing nodes, thus refining their structures flexibly and intelligently.

We summarize the contributions of this work as follows:
\begin{itemize}[leftmargin=*, topsep=0pt]
    \item We study the semi-supervised community detection problem from a novel subgraph perspective. Different from existing seed-based methods, our solution can be regarded as first locating potential communities and then refining them. 
    \item We propose a novel framework \ours, consisting of Community Locator and Community Rewriter. The community locator is proposed to quickly locate potential communities by seeking subgraphs similar to training ones. And the community rewriter can further fine-tune those located communities. Enhanced by deep reinforcement learning, it provides actions such as adding or dropping nodes, refining community structures flexibly.

    \item We conduct experiments on real-world networks as well as hybrid networks containing various types of communities. Compared with both community detection and semi-supervised community detection baselines, our model achieves outstanding performance and remains competitive on efficiency. Moreover, our model exhibits robustness even when encountering data sparsity or network noises.
 
\end{itemize}

\section{Related Work}

\subsection{Community Detection}
A common definition of community detection is to partition graph nodes into multiple groups, where internal nodes are more similar than the external.  Some works \cite{Blondel_2008, Clauset_2004, 868688} are optimization-based methods that search a graph partition by optimizing some metrics such as modularity. Moreover, there are matrix factorization based methods \cite{Li2018CommunityDI, Wang2016SemanticCI} which learn latent representations for communities by decomposing adjacency matrices. Studies like \cite{bigclam, cesna} are generative models that infer communities by fitting the original graph. In recent years, some frameworks that combine graph representation learning and community detection have been proposed \cite{come, communitygan, vgraph, commondgi}. For example, vGraph \cite{vgraph} is a probabilistic generative model to learn community membership and node representation collaboratively. In a word, these community detection works  fail to pinpoint a particular kind of community.

\textbf{Semi-supervised community detection.} This task aims to seek the rest communities in the network with few labeled communities as training data. Existing methods \cite{bespoke, seal} usually pick seed nodes in the network by leveraging training data and then develop communities around seeds. However, they are quite sensitive to the quality of selected seeds. Besides, they have individual issues, e.g., inflexibility \cite{bespoke} and high computational overhead \cite{seal}. We skip work \cite{Jin2019GraphCN} because their semi-supervised setting is completely different from current discussion.

\subsection{Subgraph Matching}
Subgraph matching refers to the task of determining the existence of the query graph in the target graph \cite{rex2020neural}. We will discuss the development of subgraph matching methods as the implementation of community locator is inspired by subgraph matching. 
Conventional algorithms such as \cite{Ullmann1976AnAF} only focus on graph structures. Other works like \cite{AlemanMeza2005TemplateBS, submatch2004} also consider categorical node features.

\textbf{Neural network based methods. }As graph neural networks (GNN) raise a surge of interest \cite{Kipf2017SemiSupervisedCW, Hamilton2017InductiveRL, Xu2019HowPA}, GNN-based matching methods have been proposed \cite{Bai2019SimGNNAN, Li2019GraphMN, Guo2018NeuralGM, rex2020neural} and have achieved great results.  A recent breakthrough in this domain is \cite{rex2020neural} which significantly outperforms other subgraph matching baselines thanks to the introduction of order embedding. Inspired by their work, we design Community Order Embedding to capture the similarities between communities and further match candidates. Especially, our novelty lies in migrating the matching method into detection tasks.

\subsection{Graph Combinatorial Optimization with RL}
With the success of deep reinforcement learning in games \cite{RL_game}, researchers have attempted to utilize reinforcement learning (RL) techniques for the graph combinatorial optimization (GCO) problem \cite{Khalil2017LearningCO, Ma2019CombinatorialOB, seal, RG-Explainer}. S2V-DQN \cite{Khalil2017LearningCO} is the first attempt that models a GCO problem into a sequential decision problem with deep Q-learning. And in\space \cite{Ma2019CombinatorialOB} the authors propose Graph Pointer Networks to solve TSP problems. SEAL \cite{seal} adopts policy gradient to learn heuristics for generating a community. Recently, \citet{RG-Explainer}  propose a RL-based framework named RG-Explainer for generating explanations for graph neural networks. Note that rewriting a community, i.e., adjusting its members, is also a GCO problem. Thus, in this paper, we resort to RL for optimizing the community rewriter.

\section{Methodology}

Given a graph $G=(\mathcal{V},\mathcal{E},\mathbf{X})$, where $\mathcal{V}$ is the set of nodes, $\mathcal{E}$ is the set of edges, and $\mathbf{X}$ is the node feature matrix. With $m$ labeled communities $\dot{\mathcal{C}}=\{\dot{C}^1, \dot{C}^2, ..., \dot{C}^{m}\} ( \forall_{i=1}^m \dot{C}^i \subset G )$ as training data, our goal is to find the set of other similar communities $\hat{\mathcal{C}}$\space($|\hat{\mathcal{C}}| \gg |\dot{\mathcal{C}}|$) in $G$. The important notations used are listed in Table\space\ref{table:notation}.

\begin{table}[H]
  \caption{Important Notations}
  \label{table:notation}
  \begin{tabular}{c|c}
    \toprule
      Notations & Definition\\
    \midrule
    $G=(\mathcal{V},\mathcal{E},\mathbf{X})$ & Network \\
    
    $\dot{\mathcal{C}} = \{ \dot{C}^1, ..., \dot{C}^m \}$ & The set of training communities \\
    
   $ \mathcal{N}(u)$ & Neighbors of node $u$ \\ 
    $\mathbf{x}(u)$ & Raw features of node $u$ \\
    $\mathbf{x}'(u)$ & Augmented features of node $u$ \\
    $\mathbf{z}(u)$ & Embedding of node $u$ \\
    $\mathbf{z}(C^i)$ & Embedding of community $C^i$\\
    
    
    $\partial C=\cup_{u\in C}\mathcal{N}(u)\setminus C$ & Outer boundary of the community $C$\\
    
   $\mathbf{s}_t(u)$ & Representation of node $u$ at time $t$   \\

  \bottomrule
\end{tabular}
\end{table}

\begin{figure*}
    \centering
    \includegraphics[width=16.5cm]{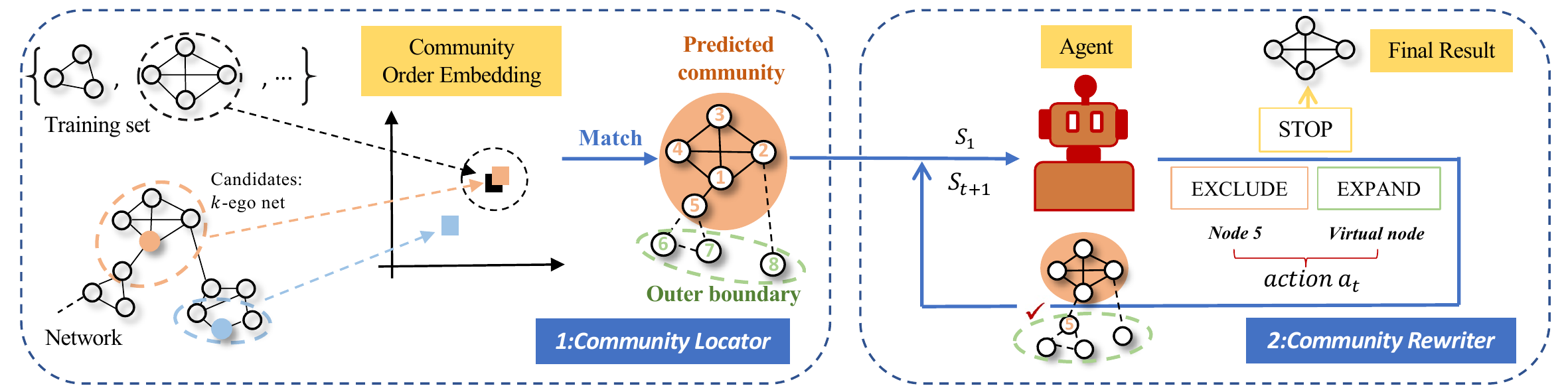}
    \caption{\ours \space inference framework. For a specified training community, suppose our goal is to detect a new community similar to it in the network. We utilize community locator for locating the best-matched candidate community. We further rewrite this candidate community by feeding it to a well-trained structure agent. After performing incremental ``EXCLUDE'' and ``EXPAND'' actions suggested by this agent, we obtain the final predicted community.}
    \label{fig:framework}
\end{figure*}

As shown in Figure \ref{fig:framework}, we first describe the inference process of \ours. It consists of 2 components, \textbf{Community Locator} and \textbf{Community Rewriter}. For a specified training community, the locator would seek its best match from all candidate communities. For efficiency, $k$-ego net\space(a central node along with its neighbors within $k$ hops) of each node in the network is regarded as a candidate. Considering that the community detected by locator may not be accurate, as its structure is fixed, we introduce Community Rewriter for further refinement. It acts like an intelligent agent that adjusts the community structures by performing actions. Specifically, it determines whether to exclude existing nodes or expand towards the outer boundary (the neighbors of nodes in this community) during each step. This iterative process is repeated until encountering ``STOP'' signal, at which point we obtain the final predicted community.

\subsection{Community Locator}
In this subsection, we explain our community locator component. Specifically, we implement community order embedding to encode each community and match candidate communities. In this way, we can quickly locate potential communities.

\subsubsection{Design goal}
We map each community into a $d$-dimensional vector and ensure that the subgraph relationship can be properly reflected in the embedding space: if community $C^a$ is a subgraph of community $C^b$, then corresponding community embedding $\mathbf{z}(C^a)$ has to be in the ``lower-left'' of embedding $\mathbf{z}(C^b)$:
\begin{equation*}
    \mathbf{z}(C^a)[i] \leq \mathbf{z}(C^b)[i],
 \;  \forall_{i=1}^{d} , \text{iff} \; C^a \subseteq C^b.
\end{equation*}

We adopt this idea of order embedding proposed in \cite{rex2020neural} to implement community order embedding. To our knowledge, migrating the matching method into detection tasks is quite novel. When the embeddings of two communities are very close to each other, we consider these two communities as isomorphic. To realize this goal, the max margin loss is utilized for training:

\small
\begin{equation*}
\mathcal{L} = \sum_{Pos}E\left(\mathbf{z}(C^a), \mathbf{z}(C^b) \right)  + \sum_{Neg}{\max \left\{0, \alpha - E\left(\mathbf{z}(C^a), \mathbf{z}(C^b) \right)\right\} } ,
\end{equation*}

\begin{equation}
  \text{where} \;   E\left(\mathbf{z}(C^a), \mathbf{z}(C^b)\right)  = \left\|\max\left\{0, \; \mathbf{z}(C^a)- \mathbf{z}(C^b)\ \right\}\right\|_2^2.
  \label{eq:distance}
\end{equation}

\normalsize

\noindent Here $Pos$ denotes the set of positive samples where community $C^a$ is a subgraph of $C^b$ in each pair $(C^a, C^b)$, $Neg$ denotes the set of negative samples, and margin $\alpha$ is a distance parameter.

\subsubsection{Identifying potential communities}
We utilize the well-trained community order embedding for identifying potential communities. In order to quickly divide the entire network into small subgraphs, we regard the $k$-ego net of each node in the network as a candidate community. As we feed these candidates to the encoder, we can get their representations $\mathbf{Z}=\{\mathbf{z}(C^1), \ ... \ ,\  \mathbf{z}(C^{|\mathcal{V}|}) \}$ where $C^i$ denotes the $k$-ego net of node $i  \in \mathcal{V}$. Similarly, training communities are encoded as $\dot{\mathbf{Z}}= \{\mathbf{z}(\dot{C}^1), ... \  , \mathbf{z}(\dot{C}^m)\}$, each of which is treated as a pattern for matching. Suppose our goal is to predict $N$ new communities, then the $n$ ($n=\frac{N}{m}$) candidate communities closest to each training one in the embedding space are considered as predicted results.

We also have an alternative method for detecting arbitrary numbers of communities in the lack of prior knowledge $N$. Specifically, we can use some threshold $\eta$ to take any community whose similarity to the closest training community is above $\eta$. Since the identified communities are ranked by the similarity measures, i.e., the distance in the latent space, we can stop when the identified community is not that similar to what we are looking for. In this way, the requirement of $N$ can be avoided.

\subsubsection{\label{sec:encoder}Graph Neural Networks based encoder}

Here we describe the concrete encoding process of communities. For a specific node $u \in \mathcal{V}$, we augment its feature as $\mathbf{x}'(u)$ = $[\mathbf{x}(u), degree(u), \text{max}(DN(u)),\\ \text{min}(DN(u)), \text{mean}(DN(u)), \text{std}(DN(u))]$ where $\mathbf{x}(u)$ is the raw features of node $u$ with default value as 1 and $DN(u)=\{degree(v)|v \in \mathcal{N}(u)\}$. Hence, the expressive power of node attributes can be boosted, which is crucial for networks without node features \cite{cai2019simple, seal}. We adopt graph neural networks (GNN) \cite{Kipf2017SemiSupervisedCW, Hamilton2017InductiveRL, Xu2019HowPA} as the basic component of the encoder $Z$.

Firstly, the original node representations undergo a linear transformation as $ \mathbf{z}^{(0)}(u) = \mathbf{W}^{1} \cdot \mathbf{x}'(u)$. Then, the encoder propagates and aggregates the information via a \textbf{$k$-layers GNN}. We study the impact of different graph neural networks in the experiments and will discuss later. Note that we ensure the number of layers is in consistent with the choice of ego net dimension $k$. Finally, we concatenate the node embeddings in previous layers and perform linear transformation again to obtain the final embedding $\mathbf{z}(u)$ of node $u$ as follows:
\begin{equation*}
\setlength{\abovedisplayskip}{3pt}
\setlength{\belowdisplayskip}{3pt}
    \mathbf{z}(u) = \mathbf{W}^{2}\cdot \textbf{CONCAT}\left(\mathbf{z}^{(0)}(u), ... \;, \mathbf{z}^{(k)}(u)\right), 
\end{equation*}

\noindent where $\mathbf{z}^{(l)}(u)$ denotes the embedding of node $u$ in the $l$-th layer and $\mathbf{W}^{j} (j=1, 2)$ are trainable parameters. For a specific community $C^i$, its embedding is calculated as $\mathbf{z}(C^i) = \sum_{u \in C^i} \mathbf{z}(u)$. 

So far, the optimization process is summarized in Algorithm\space\ref{algorithm:1}.

\begin{figure*}
    \centering
    \includegraphics[width=17cm]{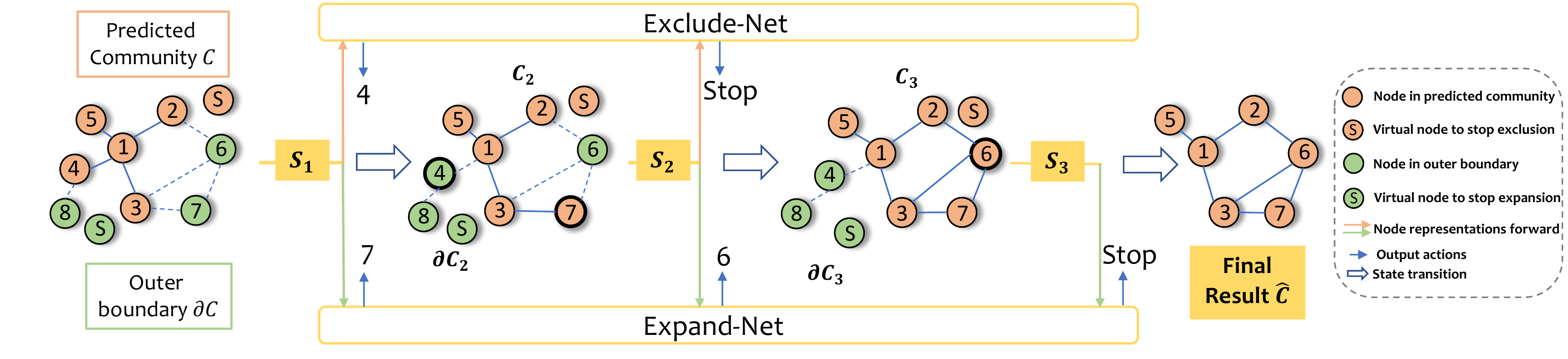}
    \caption{Community Rewriting Process. For a community $C$ detected in the first stage, we associate it with its outer boundary $\partial C$. At the $t$-th step, Exclude-Net would pick a node in $C_t$ for exclusion. Similarly, Expand-Net selects a node from $\partial C_t$ for expansion. The next state is moved to after performing both Exclude and Expand actions. If both Exclude-Net and Expand-Net have decided to ``STOP'', the refinement ends and final predicted community is generated.}
    \label{fig:rewriting}
\end{figure*}

\subsection{Community Rewriter}
In Community Locator, we make an assumption on the structure of predicted communities for efficiency while sacrificing flexibility. Therefore, Community Rewriter is further proposed to provide flexibility via intelligent refinement.

The core idea is that, given a raw predicted community $C$, we can either drop some irrelevant nodes in $C$ or absorb nodes from its outer boundary, i.e., $\partial C=\cup_{u\in C}\mathcal{N}(u)\setminus C$, to obtain a more accurate estimation. Such structural refinement can be considered as a combinatorial optimization problem, where we need to pick suitable nodes in a set to optimize the objective. Therefore, we adopt RL for automatically learning heuristics to rewrite communities. An illustrative example of rewriting process  is shown in Figure  \ref{fig:rewriting}.

\subsubsection{Task design} The refinement of a community detected by the locator is regarded as an episode. Specifically, the initial state is defined as $S_1 = C \cup \partial C$ where $C$ is a raw predicted community and $\partial C$ is its boundary. At the $t$-th step, we have the current predicted community $C_t$ and its boundary $\partial C_t$. We further pick a node $a_t^{\text{exclude}}$ in $C_t$ for exclusion as well as a node $a_t^{\text{expand}}$ from $\partial C_t$ for expansion. We would modify the community as $C_{t+1}=C_{t} \setminus \{a_t^{\text{exclude}} \} \cup \{ a_t^{\text{expand}} \}$. In this way, we move to the next state. It is worth noting that the state is the combined representations of both $C_t$ and $\partial C_t$. The action space for exclusion is $C_t$ while $\partial C_t$ serves for expansion. We also assign a reward value for rewriting.

\textbf{State.} As mentioned before, the state at timestamp $t$ is denoted as $S_t=C_t \cup \partial C_t$. For any node $u \in S_t$, we give it a node representation $\mathbf{s}_t(u)$. We design the initial node representation $\mathbf{s}_1(u)$ by augmenting its first-stage embedding $\mathbf{z}(u)$ with the information of raw community $C$:

\begin{equation*}
\small
    \mathbf{s}_1(u) = \textbf{CONCAT}\left(\mathbf{z}(u),  \mathbb{I}\{u \in C\} \right),
\end{equation*}

\noindent where $\mathbb{I}\{ \text{condition} \}$ is the indicator function that returns 1 if the condition is true, otherwise 0.

Each node could further combine information from its current neighborhood. To achieve this, we also utilize a GNN-based network $D$ parameterized by $\theta$. Thus, at the $t$-th step, the representation of node $u$ is updated as:
\small
\begin{equation*}
    D_{\theta}(\mathbf{s}_{t}(u)) = \textbf{GNN}(\mathbf{s}_{t}(u), \{\mathbf{s}_t(v) | v \in S_t \}),
\end{equation*}
\begin{equation*}
    \mathbf{s}_{t+1}(u) = \textbf{CONCAT}\left( D_{\theta}(\mathbf{s}_{t}(u) ), \mathbb{I}\{u \in C_{t+1}\} \right).
    \label{eq:node_upate}
\end{equation*}
\normalsize

\textbf{Action}. During each step, we consider adding a new node and dropping an existing node simultaneously. In other words, the action suggested at the $t$-th step is composed of both $a_t^{\text{exclude}}$ and $a_t^{\text{expand}}$. Specifically, we take $C_t$ as the action space for ``EXCLUDE'' while $\partial C_t$ serves for ``EXPAND''. 

We design respective policy networks for these two kinds of actions. They share similarities in the overall architecture: basic Multilayer Perceptron (MLP) \cite{MLP} ends with a softmax layer to measure the probability of taking a specified node:
\small
\begin{equation*}
      a^{\text{exclude}}_t = U_{\phi}(C_t) = \textbf{Softmax}\left( \textbf{MLP}_{\phi}(\{ \mathbf{s}_t(u) | u \in C_t \}) \right),
\end{equation*}
\begin{equation*}
    a^{\text{expand}}_t = P_{\psi}(\partial C_t) = \textbf{Softmax} \left( \textbf{MLP}_{\psi}(\{\mathbf{s}_t(v) | v \in \partial C_t \}) \right),
\end{equation*}
\normalsize

\noindent where $\phi$ is the trainable parameter of Exclude-Net $U$ and $\psi$ is the trainable parameter of Expand-Net $P$.

We add a special ``STOP'' action into the action space to learn stopping criteria. Concretely, we construct a \textbf{virtual node} to represent the stop signal. It is an isolated node with all-zero features. If Exclude-Net selects the virtual node, we will stop excluding and merely expand towards the outer boundary. Similarly, we do not expand when the virtual node is chosen for expansion. If both exclusion and expansion end, so does the refinement process.

\textbf{Reward.} Since our rewriting goal is to obtain a more accurate community estimation via incremental actions, we directly take one of the evaluation metrics for reward computation, i.e., F1 score. Given the suggested action $a_t$ at the $t$-th step, we define the reward $r_t$ at that time as the difference of F1 score brought by $a_t$.

\subsubsection{Optimization} We learn the rewriting agent $A=\{ \phi, \psi, \theta \}$ via \textbf{policy gradient} \cite{Sutton1999PolicyGM, Kakade2001ANP}. Since those three parameters share the similar process of being updated, we take the Exclude-Net $U$ parameterized by $\phi$ as an example to illustrate the optimization.

Given a community $C$ detected in the first stage, we can form an exclusion trajectory $\tau=\{ S_1, a_1, S_2, a_2, ..., S_T, a_T, S_{T+1} \}$ \space(we omit the superscript ``exclude'' for brevity here). Then the reward $r_t$ obtained by performing the Exclude action $a_t$ is computed as:
\begin{equation}
\small
\label{eq:reward}
    r_t = \delta(C_{t+1}, \dot{C}^i) - \delta(C_t, \dot{C}^i),
\end{equation}

\noindent where $\delta$ denotes F1 score and $\dot{C}^i$ denotes the corresponding ground-truth community. Following the same way, we further calculate rewards for all Exclude actions $a_t (\forall_{t=1}^{T} )$ in $\tau$. Finally, we update  $\text{Exclude-Net}_\phi$ according to policy gradient:

\begin{equation}
  \small
    \phi \leftarrow \phi + lr  \sum_{t=1}^{T} \nabla_{\phi} \log{U_{\phi}(a_t | S_t)} \cdot r_t,
    \label{eq:vpg}
\end{equation}
\noindent where $lr$ stands for the learning rate of optimizing Exclude-Net $U$ parameterized by $\phi$.

To realize the above optimization objective, a lot of training samples are constructed. Specifically, the generation of a sample follows this way: firstly, we randomly pick a node $u$ from a training community $\dot{C}^i$. Then, its $k$-ego net $C^{u}$ as well as corresponding boundary $\partial C^{u}$ are extracted. Repeatedly, the set of training samples is constructed as $\mathcal{D} = \{ (C^u \cup \partial C^{u}, \dot{C}^i) \}$ where $C^u$ is a $k$-ego net with $u$ coming from some training community $\dot{C}^i \in \dot{\mathcal{C}}$. Notably, we fix the structure of these training samples as \textbf{$k$-ego net} to simulate the communities detected by the community locator and $\dot{C}^i$ is utilized for calculating the rewards.

For a training sample $(C^{u} \cup \partial C^{u}, \dot{C}^i)$, the agent will produce a trajectory $\tau = \{(S_t, a_t, r_t, S_{t+1})\}$ where $a_t$ is composed of $a_t^{\text{exclude}}$ and $a_t^{\text{expand}}$. We calculate the reward $r_t$ associated with $\dot{C}^i$ according to Equation~\ref{eq:reward}. The next state $S_{t+1}$ is generated by taking the action $a_t^{\text{expand}}$ and $a_t^{\text{exclude}}$ from the current state $S_{t}$. When the Expand or Exclude action selects the virtual node, this kind of action will stop. If both expansion and exclusion are stopped, the episode ends and we obtain a complete trajectory $\tau$. Based on the collected trajectory $\tau$ for each training sample, we update parameters in the agent. The detailed process is described in Algorithm\space\ref{algorithm:3}.

\subsection{Summary}
We summarize the \ours \space framework as shown in Algorithm\space\ref{algorithm:smrc}. Recall that with $m$ labeled communities as training data, our goal is to detect $N$ new communities in the network. 

Firstly, we train the community locator by leveraging known communities. Then we take each training community as a pattern for matching $n$ closest candidate communities in the embedding space ($n=\frac{N}{m}$). Actually, the $k$-ego net of each node in the network serves as a candidate. After matching, we can get $N$ raw predicted communities. Next, we train the community rewriter. For each community detected in the first stage, it is fed to this well-trained agent and refined into a new community. Finally, we obtain $N$ modified communities as final results.

\section{Experiments}

In this section, we conduct extensive experiments to verify both the effectiveness and efficiency of \ours \space on multiple datasets. We also compare the robustness of existing semi-supervised community detection algorithms under different conditions. Due to the space limitation, we move the implementation details and parameters study to Appendix.

\subsection{Experiment Setup}
\textbf{Evaluation metrics.} 
For networks with ground-truth communities, the most used evaluation metrics are bi-matching \textbf{F1 score} and \textbf{Jaccard score}\space \cite{bespoke, seal, communitygan, metrics_survey}.
Given  $M$ ground truth communities $\{\dot{C}^{j}\}$ and $N$ generated communities $\{\hat{C}^{i}\}$, we compute scores as:

\begin{equation}
\small
\frac{1}{2} \left(\frac{1}{N} \sum_{i}{\max_{j} \delta(\hat{C}^{i}, \dot{C}^{j})} + \frac{1}{M} \sum_{j}{\max_{i} \delta(\hat{C}^{i}, \dot{C}^{j})} \right) ,
\label{eq:metrics}
\end{equation}

\noindent where $\delta$ can be F1 or Jaccard function. Besides, we use the overlapping normalized mutual information (\textbf{ONMI}) \cite{onmi} as a supplementary metric, which is the overlapping-version of NMI score. For more information of ONMI, please refer to \cite{onmi}. 

 \textbf{Datasets.} To comprehensively assess the effectiveness of our model, we conduct experiments both on \textbf{single datasets} (a network with partially labeled communities) and \textbf{hybrid datasets} (combination of multiple different single datasets). 
 
 \noindent \textbf{Single datasets.} We choose three common real-world networks containing overlapping communities from SNAP\footnote{http://snap.stanford.edu/data/} (Amazon, DBLP, and Livejournal). Note that these networks are partially labeled, i.e., most nodes do not belong to any community. Thus, we can view that there are other types of communities in the networks, and our targeted communities are the labeled ones. 
 
 \noindent \textbf{Hybrid datasets. } We create hybrid networks by combining two different networks following related works \cite{seal}. For example, we stack Amazon and DBLP by randomly adding 5,000 cross-networks links between two graphs, resulting in a bigger network. Since communities in different networks exhibit different features \cite{Yang2012DefiningAE}, we obtain a single network with various types of communities in this way. Similarly, we combine DBLP and Livejournal by adding 10,000 cross-networks links. These two networks are named ``Amazon+DBLP'' and ``DBLP+Livejournal'', respectively. Note that we only add some cross-networks links, so the internal connectivity between communities will not be disturbed.

 The statistics of all datasets are shown in Table\space\ref{table:datasets}. For each single dataset, we use 90 communities as the training set, 10 as the validation set, and the rest as the test set. As to hybrid datasets, we aim to pinpoint one kind of community. For example, on Amazon+DBLP, we would take 90 communities from Amazon for training, 10 for validation, and the remaining communities from Amazon serve for testing.

 \begin{table}[t]
 \small
  \caption{Statistics of datasets. $C_{Max}$ denotes the largest community size while $C_{Avg}$ denotes the average community size.}
  \label{table:datasets}
  \begin{tabular}{ccccccc}
    \toprule
      &$\#N$& $\#E $& $\#C$&  $C_{Max}$ & $C_{Avg}$\\
    \midrule
    Amazon & 6,926 & 17,893 & 1,000 & 30 & 9.38 \\
    DBLP & 37,020 & 149,501 & 1,000 & 16 & 8.37 \\
    Livejournal & 69,860 & 911,179 & 1,000 & 30 & 13.00\\
    \midrule
    Amazon+DBLP & 43,946 & 172,394 & 2,000 & 30 & 8.88 \\
    DBLP+Livejournal & 106,880 & 1,070,680 & 2,000 & 30 & 10.69\\
  \bottomrule
\end{tabular}
\end{table}

\begin{table*}[htp]
   \small
    \centering
    \caption{Summary of the performance in comparison with baselines. N/A means the model fails to converge in 2 days. We report the results of \ours \space with $k$=1 on DBLP while $k$=2 on all other datasets.}
    \label{tab:experiment_result}
    \begin{tabular}{c|cccccc|ccc}
        \toprule
         & Dataset & BigClam & BigClam-A & ComE & CommunityGAN & vGraph & Bespoke & SEAL & \ours \\
        \midrule
        \multirow{7}{*}{\textbf{F1}} & Amazon & 0.6885 & 0.6562 & 0.6569 & 0.6701 & 0.6895 & 0.5193 & \underline{0.7252} & \textbf{0.7730} \\ 
        
        & DBLP & 0.3217 & 0.3242 & N/A & \underline{0.3541} & 0.1134 & 0.2956 & 0.2914 & \textbf{0.3835} \\ 
        
        & Livejournal & 0.3917 & 0.3934 & N/A & 0.4067 & 0.0429 & 0.1706 & \underline{0.4638} & \textbf{0.4950} \\

        & $\text{Amazon}^{*}\text{DBLP}$ & 0.1759 & 0.1745 & N/A & 0.0204 & 0.0769 & 0.0641 & \underline{0.2733} & \textbf{0.3988} \\
        
        & $\text{DBLP}^{*}\text{Amazon}$ & 0.2363 & 0.2346 & N/A & 0.0764 & 0.1002 & \underline{0.2464} & 0.1317 & \textbf{0.2901} \\ 
        
        & $\text{DBLP}^{*}\text{Livejournal}$ & 0.0909 & 0.0859 &N/A  & 0.0251 & 0.0131 & 0.0817 & \underline{0.1906} & \textbf{0.2480} \\ 
         
        & $\text{Livejournal}^{*}\text{DBLP}$ & 0.2183 & 0.2139 & N/A & 0.0142 & 0.0206 & 0.1893 & \underline{0.2291} & \textbf{0.2894}\\
        
        \midrule
        
        \multirow{7}{*}{\textbf{Jaccard}} & Amazon & 0.5874 & 0.5623 & 0.5691 & 0.6045 & 0.5721 & 0.4415 & \underline{0.6792} & \textbf{0.6827} \\
        
        & DBLP & 0.2186 & 0.2203 & N/A & \underline{0.2830} & 0.0645 & 0.2593 & 0.2143 & \textbf{0.3132} \\
        
        & Livejournal & 0.3102 & 0.3076 & N/A & 0.3183 & 0.0222 & 0.1324 & \underline{0.3795} & \textbf{0.4027} \\

        & $\text{Amazon}^{*}\text{DBLP}$ & 0.1102 & 0.1095 & N/A & 0.0109 & 0.0421 & 0.0488 & \underline{0.2419} & \textbf{0.3241} \\
        
        & $\text{DBLP}^{*}\text{Amazon}$ & 0.1485 & 0.1478 & N/A & 0.0610 & 0.0555 & \underline{0.2135} & 0.0879 & \textbf{0.2166}\\
        
        & $\text{DBLP}^{*}\text{Livejournal}$ & 0.0523 & 0.0485 & N/A & 0.0120 & 0.0066 & 0.0756 & \underline{0.1485} & \textbf{0.1893}\\
        
        & $\text{Livejournal}^{*}\text{DBLP}$ & 0.1505 & 0.1464 & N/A & 0.0097 & 0.0105 & 0.1503 & \underline{0.1907} & \textbf{0.2308}\\
        
        \midrule 
        
        \multirow{7}{*}{\textbf{ONMI}} & Amazon & 0.5865 & 0.5625 & 0.5570 & 0.6040 & 0.5532 & 0.4129 & \underline{0.6862} & \textbf{0.7015} \\
        
        & DBLP & 0.1113 & 0.1110 & N/A & 0.2324 & 0.0020 & \underline{0.2347} & 0.1603 & \textbf{0.2600} \\ 
        
        & Livejournal & 0.2696 & 0.2641 & N/A & 0.3171 & <1e-4 & 0.1024 & \underline{0.3695} & \textbf{0.3703} \\ 
        
        & $\text{Amazon}^{*}\text{DBLP}$ & 0.0305 & 0.0334 & N/A & <1e-4 & < 1e-4 & 0.0364 & \underline{0.2475} & \textbf{0.3126} \\ 
        
        & $\text{DBLP}^{*}\text{Amazon}$ & 0.0471 & 0.0477 & N/A & 0.0523 & <1e-4 & \textbf{0.1780} & 0.0380 & 0.1566 \\
        
        & $\text{DBLP}^{*}\text{Livejournal}$ & 0.0113 & 0.0065 & N/A & <1e-4 & <1e-4 & 0.0723 & \underline{0.1155} & \textbf{0.1331} \\
        
        & $\text{Livejournal}^{*}\text{DBLP}$ & 0.0858 & 0.0795 & N/A & 0.0053 & <1e-4 & 0.1248 & \underline{0.1906} & \textbf{0.2012}\\

        \bottomrule
    \end{tabular}
\end{table*}

\textbf{Compared baselines.} We compare \ours \space with the following methods: (1) BigClam\space \cite{bigclam} and (2) its assisted version BigClam-A, (3) ComE\space \cite{come}, (4) CommunityGAN\space \cite{communitygan}, (5) vGraph \cite{vgraph}, (6) Bespoke\space \cite{bespoke} and (7) SEAL\space \cite{seal}. Methods (1)-(5) are strong community detection baselines while (6)-(7) are semi-supervised methods requiring training communities. Other simple baselines like GCN+K-Means have been shown inferior performance\space \cite{come, commondgi}, and thus we skip those methods. We limit the numbers of their outputs to \textbf{1000} communities. For methods (1)-(5), we filter detected communities who have more than 50\% overlaps with communities in training/validation sets as SEAL \cite{seal} does.
The data pre-processing steps and comparing methods are detailed in Appendix \ref{sec:preprocessing} and \ref{sec:baseline}, respectively.

\subsection{Overall Performance}
Experimental results are shown in Table\space\ref{tab:experiment_result}. For hybrid dataset ``Amazon+DBLP'', we conduct experiments that utilize 100 communities from DBLP as prior knowledge to detect the remaining DBLP communities in the combined network, as well as utilizing 100 communities from Amazon to detect the remaining Amazon communities. They are denoted as DBLP$^{*}$Amazon and Amazon$^{*}$DBLP, respectively. Similarly, we conduct experiments named DBLP$^{*}$Livejournal and Livejournal$^{*}$DBLP. From Table \ref{tab:experiment_result}, we find that: 

\begin{itemize}[leftmargin=*, topsep=0pt]
    \item \ours \space achieves noticeable improvements on almost all datasets compared with all the baselines, demonstrating its exceptional performance. The improvements on hybrid datasets are more significant, indicating its superiority in pinpointing the targeted community.

    \item Community detection algorithms are shown their unsuitability in targeting a specific kind of community, as they perform poorly on hybrid datasets. For example, CommunityGAN \cite{communitygan} is the best baseline model on DBLP while its performance degrades dramatically on all hybrid datasets. CommunityGAN learns node-community membership matrix and assigns each node into some community. vGraph \cite{vgraph} also assumes that each node belongs to multiple communities. These approaches are more like clustering all nodes in the network rather than locating some specific type of communities. On hybrid datasets, assigning total 106,880 nodes into 1000 clusters could generate irrelevant communities on extremely large scale, resulting in inferior performance. 
    
    \item Semi-supervised community detection models (\ours, SEAL \cite{seal}, and Bespoke \cite{bespoke}) gain better predicted results on both kinds of datasets generally. For example, SEAL is the best baseline model on most datasets. 
    
    \item Bespoke performs well on DBLP, because the community structures on this dataset is closest to its assumption, 1-ego net. When other datasets do not conform to this structural assumption, the performance degrades. This also exposes the inflexibility and poor generalization of Bespoke.
    
\end{itemize}

\begin{table*}[htp]
\small 
\caption{Ablation Study on ONMI score. ``$k$-ego subgraph'' is generated by randomly selecting some subgraphs in the form of $k$-ego net. ``Locator'' denotes raw communities detected by the community locator. Note that ``\ours'' are obtained via rewriting those communities detected by the locator.}
    \label{tab:ablation}
    \centering
    \begin{tabular}{c|ccccccc}
    \toprule
    & Amazon & DBLP & Livejournal & Amazon$^{*}$DBLP & DBLP$^{*}$Amazon & DBLP$^{*}$Livejournal & Livejournal$^{*}$DBLP \\ 
   \midrule
    $k$-ego subgraph &  0.4323 & 0.1112 & 0.1140 & 0.0632 & 0.0855 & 0.0365 & 0.0726 \\ 
    Locator & 0.6586 & 0.2585 & 0.3592 & 0.3088 & 0.1546 & 0.1322 & 0.1964 \\
    CLARE & \textbf{0.7015} & \textbf{0.2600} & \textbf{0.3703} & \textbf{0.3126} & \textbf{0.1566} & \textbf{0.1331} & \textbf{0.2012}\\
    
    \bottomrule
    \end{tabular}
\end{table*}

\subsection{Ablation Study}
To evaluate the effectiveness of both components of \ours, we conduct ablation experiments. The ONMI results are shown in Table\space\ref{tab:ablation}. Due to space limitation, we omit the results of F1 score and Jaccard score, which show similar trends with ONMI.
    
\textbf{Community Locator}. We report the communities detected by the community locator as ``Locator'' in Table \ref{tab:ablation}. For an intuitive comparison, we also generate the same number of random subgraphs in the form of $k$-ego net. We can see that the locator already obtains excellent performance as the improvement compared with random subgraphs is significant. Notably, solely locator has outperformed most baselines, showing the effectiveness of locating targeted communities from the matching approach.

\textbf{Community Rewriter}. By comparing the results between ``Locator'' and ``CLARE'', it is clear that the introduction of rewriter can obtain better performance. The improvements on DBLP related datasets are relatively marginal. This is because the community structures on DBLP are very close to $1$-ego net form, it only takes few actions to rewrite.

\begin{figure}[t]
    \centering
    \includegraphics[width=7cm]{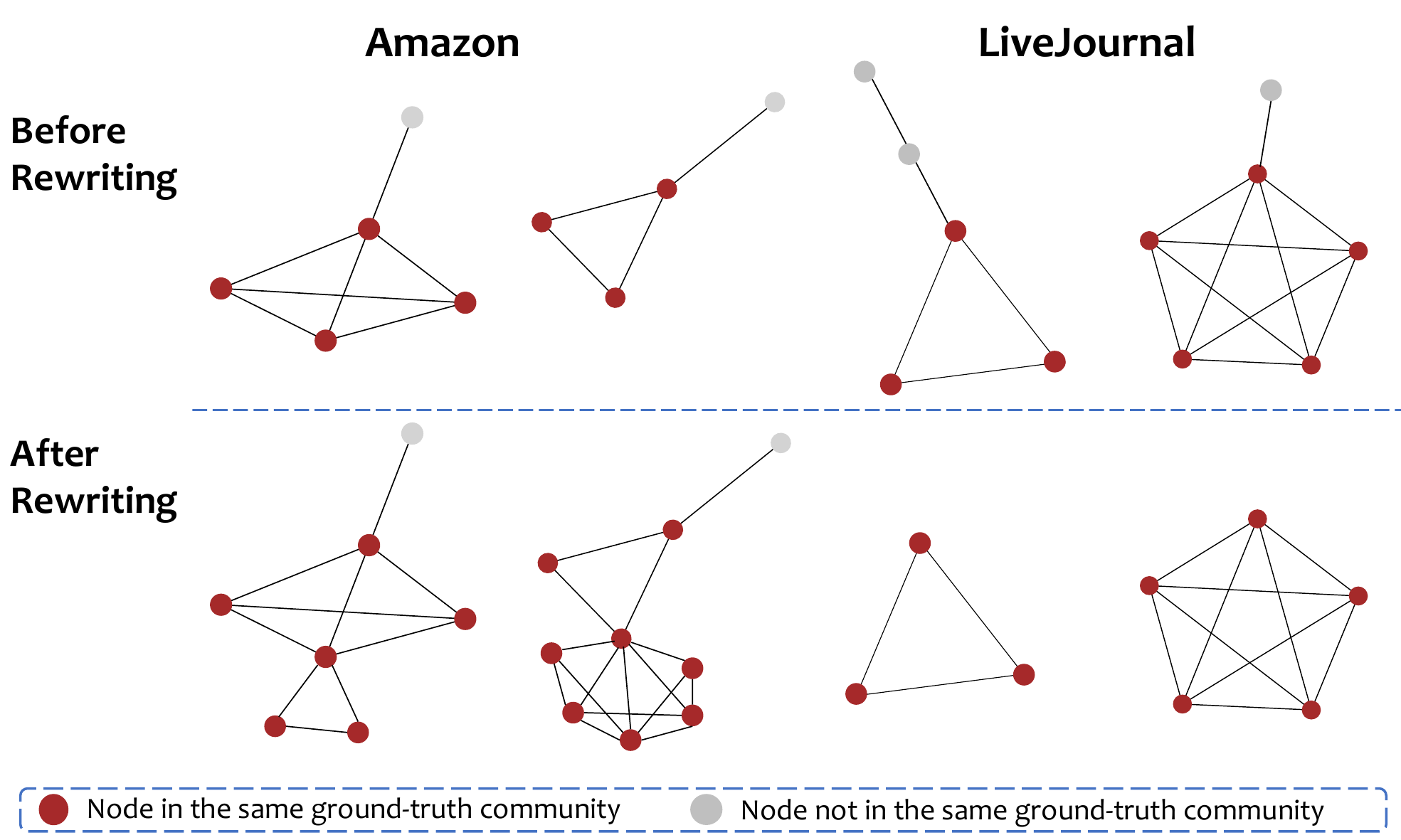} 
    \caption{Case study of the community rewriter. On Amazon, many undetected nodes can be correctly absorbed while irrelevant nodes are correctly removed on Livejournal. }
    \label{fig:cr}
    \vspace{-0.3cm}
\end{figure}

We conduct a case study to see how rewriter works as shown in Figure \ref{fig:cr}. It learns quite different heuristics for different networks, demonstrating its adaptability and flexibility. For example, on Amazon, many nodes in the same ground-truth community are not detected in the first stage, but the rewriter can intelligently absorb them. Besides, on Livejournal, though raw predicted results include few irrelevant nodes\space(not in the same ground-truth community with others), many of them can be correctly eliminated.

\subsection{Efficiency Study}

\par
\begin{figure}[t]
    \centering
    \includegraphics[width=7cm]{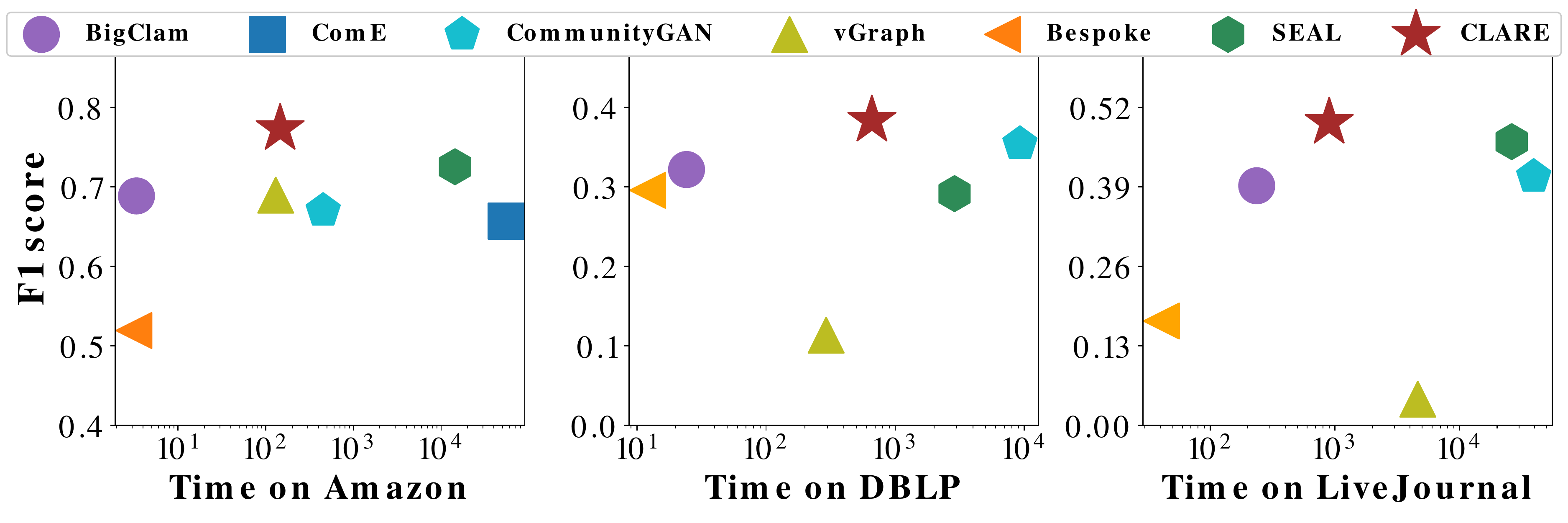}
    \centering
    \includegraphics[width=9cm]{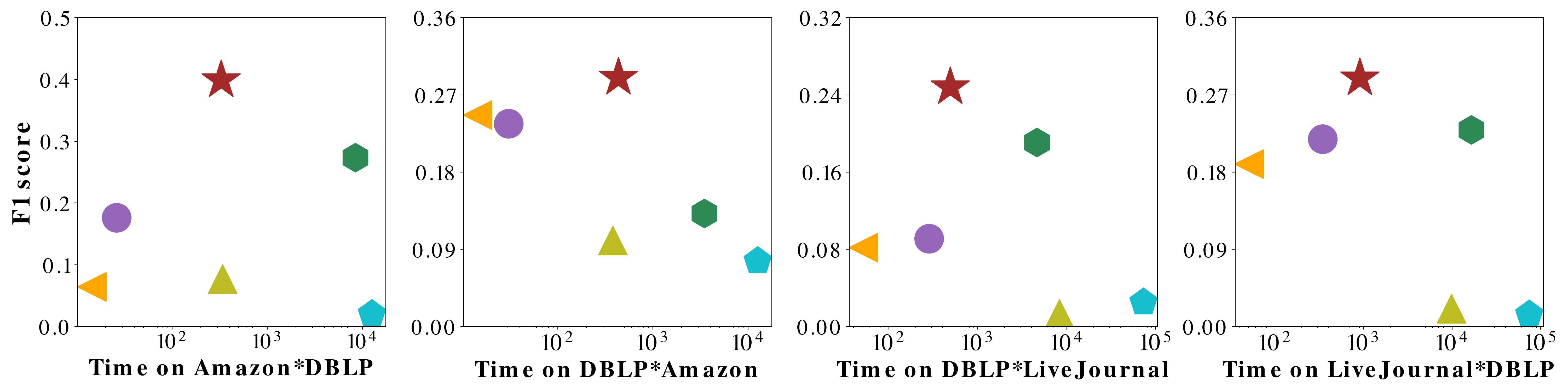}
    \caption{Efficiency comparison with all baselines.}
    \label{fig:time_comp}
\end{figure}

 We evaluate the efficiency of \ours \space by directly comparing the total running time (training plus testing) with all baselines. In this evaluation, all parameters for baselines are set following their original papers. Figure \ref{fig:time_comp} illustrates the performance (F1 score) and running time (second). Since ComE fails to converge in two days on all datasets except Amazon, we do not show it on those datasets. 

Notably, the running time of \ours \space is consistently competitive. Even on the largest dataset with totally 106,800 nodes, it takes \ours \space only about 1000 seconds. Simultaneously, its performance (F1 score) beats that of other quicker models.

\subsection{Discussions}
In this section, we compare the robustness of existing semi-supervised community detection algorithms under different conditions: (1) different numbers of training communities; (2) different levels of network noises, i.e., different numbers of cross-networks links.

\begin{figure}[t]
   \centering  
   \subfigure[Number of training communities]{   
   \begin{minipage}{4cm}
      \centering   
      \includegraphics[width=4cm]{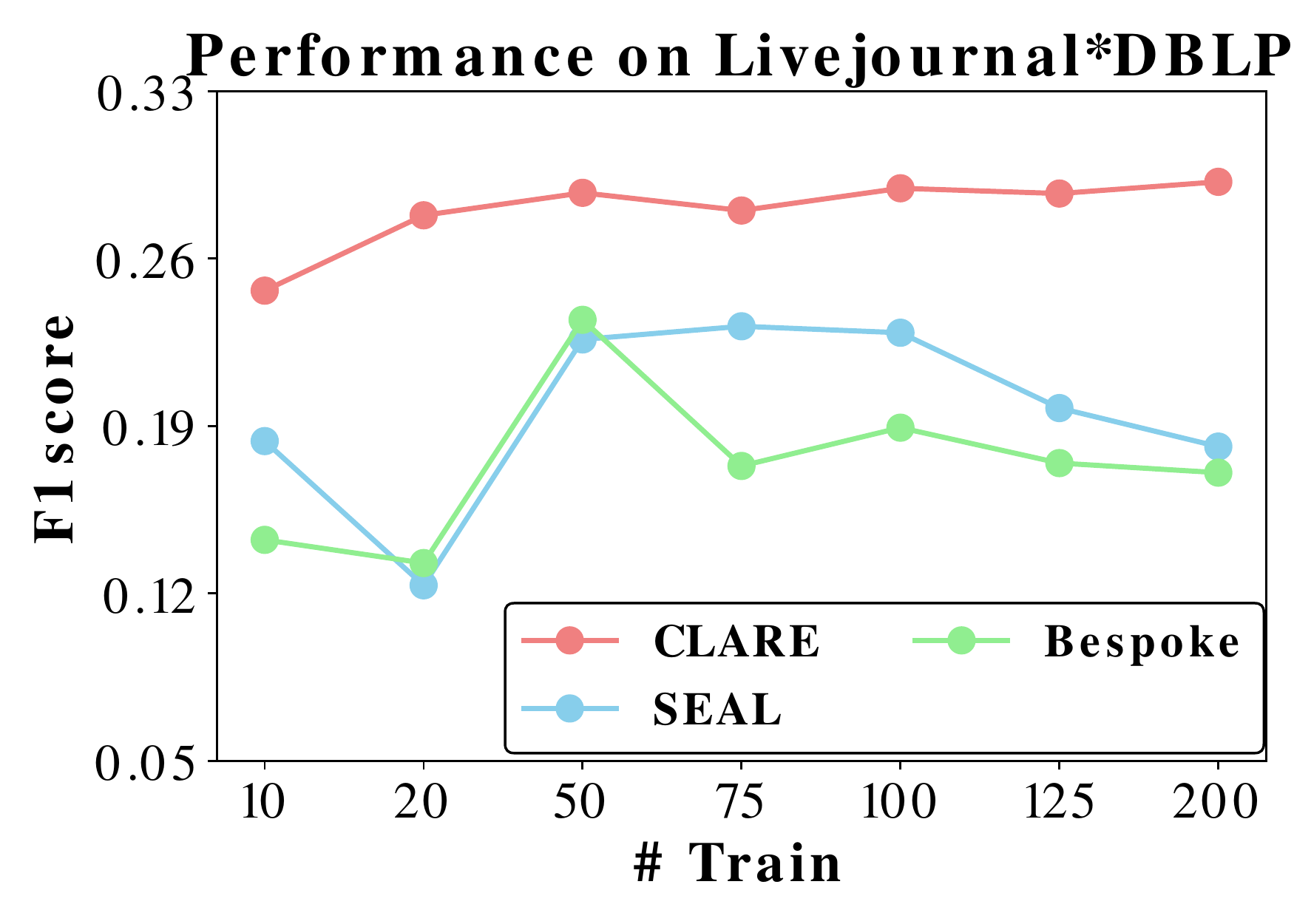}  
      \label{fig:trainsize}
   \end{minipage}
}
   \subfigure[Number of cross-networks links]{
    \begin{minipage}{4cm}
       \centering    
       \includegraphics[width=4cm]{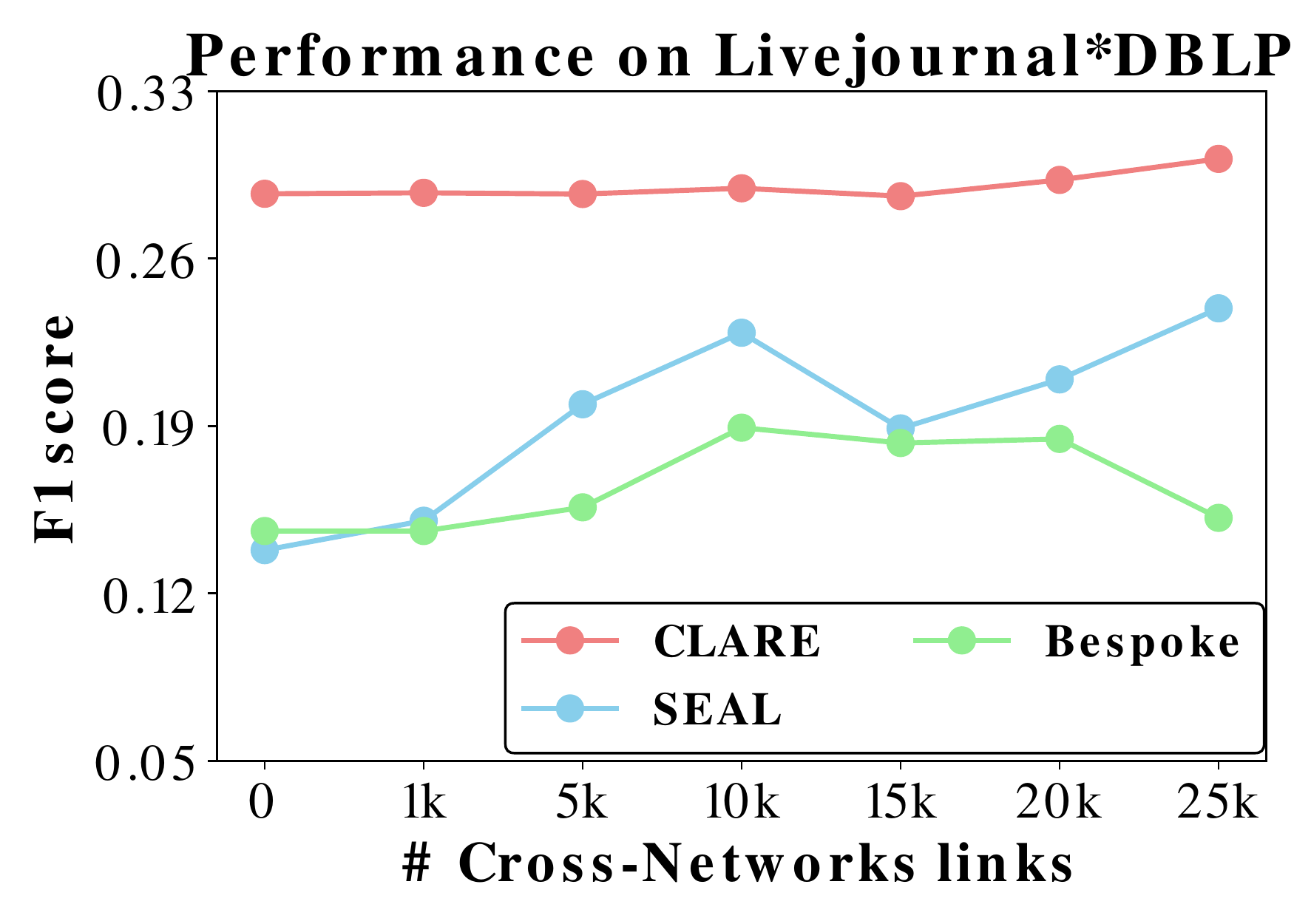}
      \label{fig:hybrid_edges}
    \end{minipage}
   }
\caption{The performance trend of semi-supervised community detection algorithms under different conditions.}    
\label{fig:disscussion}   
\end{figure}

\subsubsection{Different numbers of training communities} With the number of training communities (Livejournal) ranging from 10 to 200 progressively, we compare the performance of Bespoke, SEAL, and \ours \space on Livejournal*DBLP. As shown in Figure \ref{fig:trainsize}, we find that:

\begin{itemize}[leftmargin=*, topsep=10pt]
    \item \ours \space can learn from training data progressively. With the number of training communities increasing, the performance of \ours \space firstly undergoes improvement and then remains stable. Actually, richer subgraph patterns and rewriting patterns can be sampled from increasing training communities, resulting in more accurate results.
    
    \item On the contrary, Bespoke and SEAL show fluctuations on performance. This is because their performance relies heavily on the seed selector, which is vulnerable to limited training data.
\end{itemize}

\subsubsection{Different numbers of cross-networks links} Recall that we create hybrid datasets via adding cross-networks links. We view different numbers of added links as different levels of network noises. Because with the number of added links increasing, there will be some new densely connected subgraphs, promising to be mis-detected. We set the number of cross-networks links ranging from  0 to 25,000 and report the results in Figure \ref{fig:hybrid_edges}. We can see \ours \space is quite robust while SEAL and Bespoke are rather volatile. This also indicates that these seed-based methods are susceptible to interference from network noises.

\subsection{Application on the attributed graphs}

Considering that our experimental networks are non-attributed graphs, we supplement an experiment to show that our model can also achieve good performance on the attributed graphs. We use the Facebook dataset in SEAL \cite{seal}, consisting of 3,622 nodes, 76,596 edges, 317 node attributes, and 130 communities. We follow the same experimental settings with SEAL. 

From Table \ref{tab:facebook}, we can see that \ours \space outperforms SEAL even without node attributes and performs much better when attributes are available. It indicates that \ours \space not only captures the structural features of training communities but also effectively utilizes node attributes to obtain more accurate predicted results.

\begin{table}[htp]
\small
    \centering
    \caption{Experimental results on attributed network Facebook. \ours \space w/o attr. means \ours \space without attributes.}
    \label{tab:facebook}
    \begin{tabular}{c|c|ccc}
    \toprule
    & SEAL & \ours \space w/o attr. & \ours & \textbf{Improv} \\
    \midrule
       F1  & 0.3402 & 0.3829 & \textbf{0.4126} & 21.28\% \\
       
    Jaccard  & 0.2491 & 0.2815 & \textbf{0.3047} & 22.32\% \\
    \bottomrule
    \end{tabular}
\end{table}

\section{Conclusion}
In this paper, we study the semi-supervised community detection problem from a new subgraph perspective. We propose \ours \space where the community locator can quickly locate communities and the community rewriter further refines their structures. Specifically, we formulate the structural refinement as a graph combinatorial optimization based on RL. Experiments on real-world datasets prove both the effectiveness and efficiency of our proposal. As for future work, we will try other RL-based optimization methods to further improve the effectiveness of rewriting.

\section*{Acknowledgments}

\par This work is funded in part by the National Natural Science Foundation of China Projects No. U1936213 and No. U1636207. This work is also supported in part by NSF under grants III-1763325, III-1909323,  III-2106758, and SaTC-1930941.


\bibliographystyle{ReferenceFormat/ACM-Reference-Format}
\bibliography{reference}

\appendix
\clearpage

\section{REPRODUCIBILITY}
  We release \ours \space at \href{https://github.com/FDUDSDE/KDD2022CLARE}{https://github.com/FDUDSDE/KDD2022CLARE}. We implement \ours \space in Pytorch\footnote{https://pytorch.org/}, PyG\footnote{https://pytorch-geometric.readthedocs.io/}, and DeepSNAP\footnote{https://snap.stanford.edu/deepsnap/}. All experiments are conducted on a single NVIDIA Tesla V100 SXM2 with 32G memory.

\subsection{\label{sec:preprocessing}Data Pre-processing}
\par We use the networks with ground-truth communities (Amazon, DBLP, and Livejournal) provided by SNAP. For conducting experiments, we perform the following pre-processing:
 \par (1) We omit communities whose size are beyond the 90-th percentile. For example, the largest community in DBLP contains 7,556 nodes, while the 90-th percentile is only 16. By doing so, we can exclude outliers. 
 
 \par (2) Furthermore, we randomly select 1000 communities from these retrieved ones. This number is a trade-off between maintaining a relatively large network and being scalable on most baselines. Note that ComE \cite{come}, CommunityGAN \cite{communitygan}, and vGraph \cite{vgraph} mainly use networks with thousands of nodes in their original papers. They can hardly execute on networks consisting of total communities due to huge memory utilization.
 
 \par (3) For each dataset, we extract a subgraph that contains only the nodes in communities and their corresponding outer boundaries. In this way, we obtain the final datasets for experiments.
 \par For hybrid datasets, given that Amazon and Livejournal are significantly different in size (almost 1:10 in scale), we skip merging these two datasets and just consider Amazon+DBLP and \\DBLP+Livejournal combinations.

\subsection{\label{sec:baseline}Comparing methods}

During experiments, we consider both community detection and semi-supervised community detection strong baselines. 

\noindent \textbf{Community detection algorithms}:

\begin{itemize}[leftmargin=*, topsep=2pt]
    \item  \textbf{BigClam} \cite{bigclam}: This is a strong baseline for overlapping community detection based on matrix factorization. We also consider an assisted version of BigClam following the work\space \cite{bespoke}, denoted by \textbf{BigClam-A}.
    
    \item \textbf{ComE} \cite{come}: This is a framework that jointly optimizes community embedding, community detection, and node embedding.
    
    \item \textbf{CommunityGAN} \cite{communitygan}: This is a method that extends the generative model of BigClam from edge level to motif level. 
    
    \item \textbf{vGraph} \cite{vgraph}: This is a probabilistic generative model to learn community membership and node representation collaboratively.
    
\end{itemize}

\noindent \textbf{Semi-supervised community detection algorithms}:

\begin{itemize}[leftmargin=*, topsep=2pt]
    \item \textbf{Bespoke} \cite{bespoke}: This is a semi-supervised community detection algorithm based on structure and size information. 
    
    \item \textbf{SEAL} \cite{seal}: This is the start-of-the-art semi-supervised community detection algorithm that aims to learn heuristics for community detection based on GAN. 
\end{itemize}

Executable file for BigClam is from SNAP. Codes for ComE, CommunityGAN, vGraph, Bespoke, SEAL are provided by the authors.  

\par ComE \cite{come} can not be converged within 2 days on most datasets, so we report N/A. CommunityGAN \cite{communitygan} and vGraph \cite{vgraph} suffer from high memory utilization. So we employ the mini-batch strategy with the batch size of 5000 or 15000 for optimizing vGraph. As to CommunityGAN, we replace its Adam optimizer with a SGD optimizer for reducing memory usage. In addition, we all follow their default parameters settings.

\subsection{Implementation Details}
In this section, we emphasize some details of \ours \space implementation. Hyper-parameters are summarized in Table \ref{tab:hyperparameters1}.


\textbf{Community Size:} From Table\space\ref{table:datasets} we can see some networks are rather dense and sometimes the size of $1$-ego net may exceed the maximum size of communities. Therefore, we set the maximum size for generated communities as the maximum size of training ones. 

\textbf{Outer Boundary Size: } As mentioned before, some networks are quite dense, the size of a specific outer boundary may by huge, resulting in slow convergence for the optimization of the rewriter. Thus, we fix the maximum size of outer boundary as 10.

\begin{table*}[htp]
    \small
    \caption{Comparison with different graph neural network encoders. Locator results are reported.}
    \label{tab:encoder_design}
    \centering
    \begin{tabular}{c|ccc|ccc|ccc}
    \toprule
     & \multicolumn{3}{c}{Amazon} &  \multicolumn{3}{c}{DBLP}  &\multicolumn{3}{c}{Livejournal}   \\
     & F1 & Jaccard & ONMI & F1 & Jaccard & ONMI & F1 & Jaccard & ONMI \\
    \midrule
     GCN & \textbf{0.7438} & \textbf{0.6473} & \textbf{0.6586} & 0.3819 & \textbf{0.3116} & \textbf{0.2585} & 0.4899 & 0.3953 & 0.3592 \\
    
    GIN & 0.7169 & 0.6196 & 0.6313 & \textbf{0.3841} & 0.3100 & 0.2561 & \textbf{0.4943} & \textbf{0.4004} & \textbf{0.3660} \\ 
   
    GAT & 0.7231 & 0.6235 & 0.6318 & 0.3751 & 0.3021 & 0.2446 & 0.4745 & 0.3806 & 0.3405 \\
    
    \bottomrule
    \end{tabular}
\end{table*}

\begin{table}[H]
    \small
    \centering
    \caption{Hyper-parameters in CLAER}
    \label{tab:hyperparameters1}
    \begin{tabular}{c|c|c}
    \toprule
    Component &  Hyper-parameter  & Value \\
    \midrule     
     \multirow{9}{*}{\textbf{Locator}} &  Batch size     & 32 \\
     &  Number of samples in one batch & 50 \\
     &  Number of epochs & 2 \\ 
     &  Embedding dimension & 64 \\ 
     &  $k$ \& GNN layers & Searched from \{1, 2\} \\ 
     &  Learning rate & 1e-4 \\ 
     &  Optimizer & Adam \\ 
     &  Dropout rate & 0.2 \\ 
     &  Margin $\alpha$ & 0.4 \\ 
     
     \midrule
     
    \multirow{9}{*}{\textbf{Rewriter}} &  MLP of Exclude-Net & 65-32-1 \\
   & MLP of Expand-Net & 65-32-1 \\ 
   & Embedding updater & 65-64 GIN \\ 
   & Optimizer & Adam \\ 
   & Maximum size of outer boundary & 10 \\ 
   & Epoch & 1200 \\ 
   & Episode for one epoch & 20 \\

     \bottomrule
    \end{tabular}
\end{table}

\begin{figure}[htbp]
   \centering  
   \subfigure[Performance on Amazon]{   
   \begin{minipage}{4cm}
      \centering   
      \includegraphics[width=4cm]{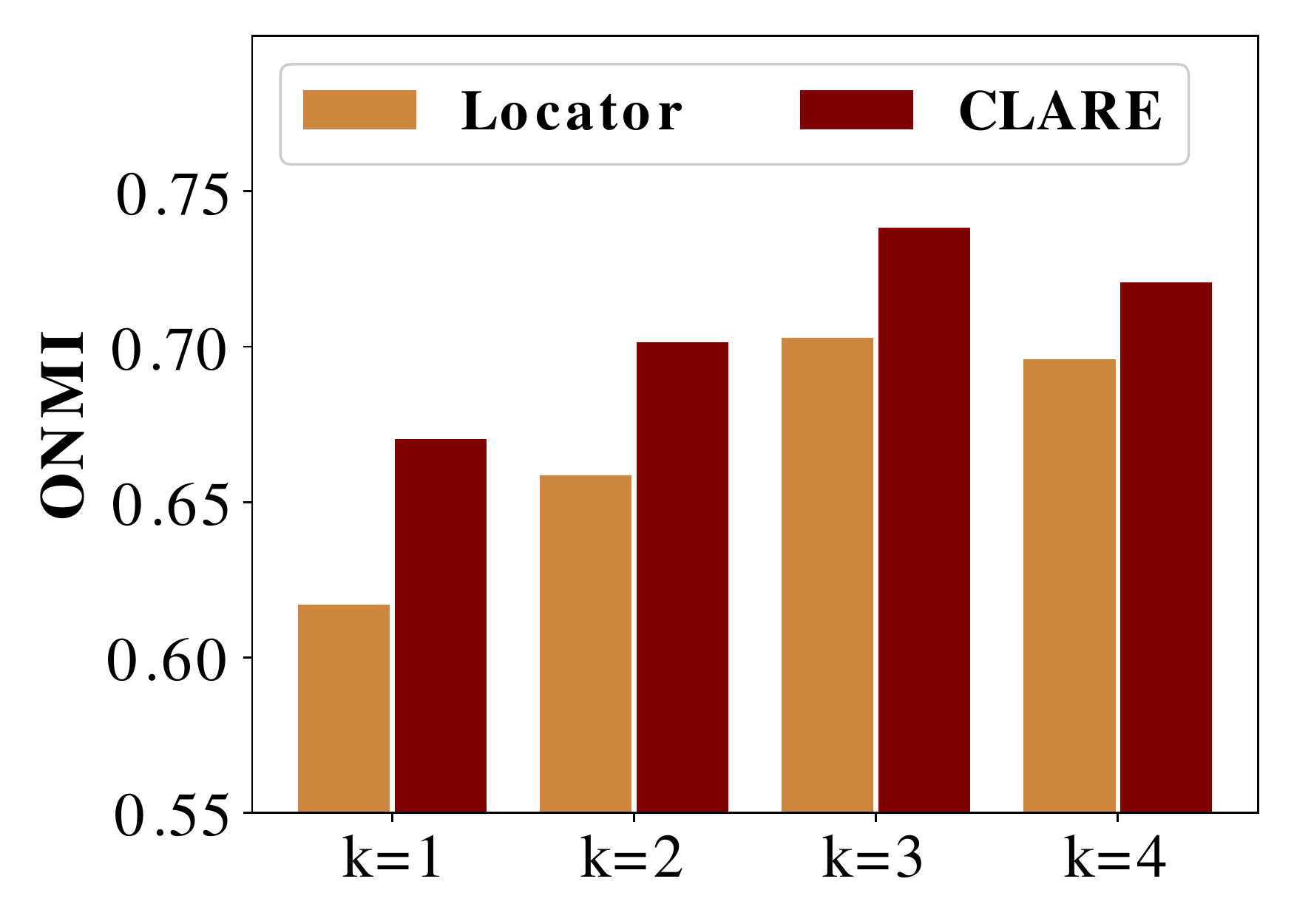}  
      \label{fig:amazon_k}
   \end{minipage}
}
   \subfigure[Performance on DBLP]{ 
    \begin{minipage}{4cm}
       \centering  
       \includegraphics[width=4cm]{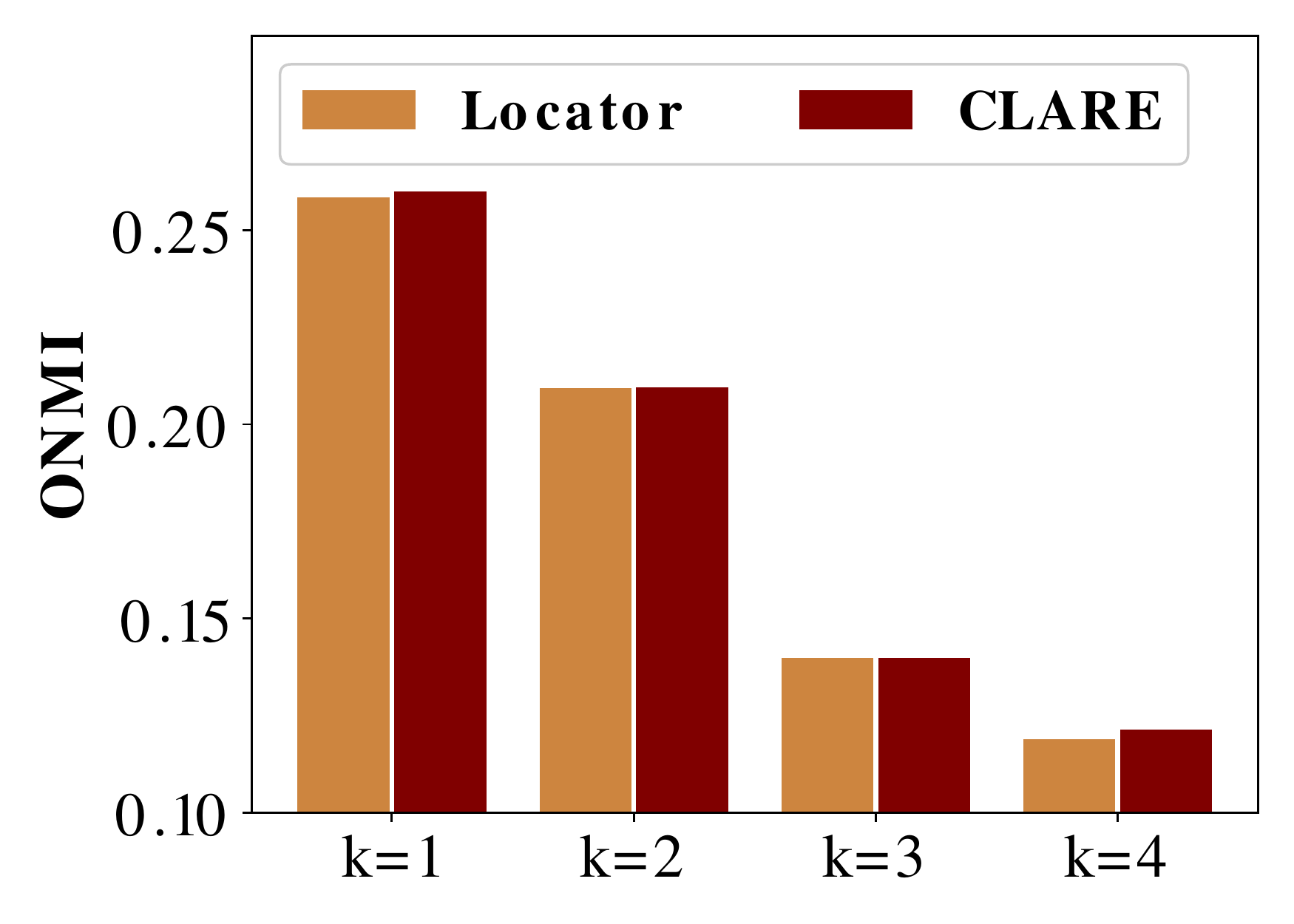}
      \label{fig:dblp_k}
    \end{minipage}
   }
  \caption{Comparison with different choices of $k$}    
\label{fig:k_choice}  
\end{figure}

\section{EXPERIMENTS}

\subsection{Design of Architectures}
\par \textbf{Design of Encoder in Community Locator: }For better architecture and performance, we conduct experiments about the design of encoder in the community locator. We choose three different graph neural networks as the encoder to learn node and community representations, including graph convolutional network (GCN) \cite{Kipf2017SemiSupervisedCW}, graph isomorphism network (GIN) \cite{Xu2019HowPA}, and graph attention network (GAT) \cite{Hamilton2017InductiveRL}. 
For the fairness of comparison, on DBLP, we fix the number of GNN layers as \textbf{1} while \textbf{2} on Amazon and Livejournal. We report Locator results in Table \ref{tab:encoder_design}. As can be observed, GCN achieves the best performance on Amazon and DBLP. Although GIN can be competitive on Livejournal, because GIN demands more training time and memory, we choose \textbf{GCN} as our encoder finally. Since our encoding objects are small subgraphs, simple graph neural networks are expressive enough \cite{subg_con, pitfall}. 

\subsection{Parameters Study}

In this part, we examine the influence of three key parameters. 

\par \textbf{Choices of $k$:} Due to communities in different datasets exhibiting distinct features \cite{Yang2012DefiningAE}, actually $k$ is an important parameter that needs to be tuned. The comparison with different choices of $k$ is depicted in Figure \ref{fig:k_choice}. The trend on different datasets varies, as Amazon undergoes performance improvement with the increase of $k$ while DBLP deteriorates. Therefore, we set $k$=1 for DBLP while $k$=2 for the remaining datasets during experiments.

\par
\begin{figure}[t]
   \centering  
   \subfigure[Different embedding dimensions]{   
   \begin{minipage}{4cm}
      \centering   
      \includegraphics[width=4cm]{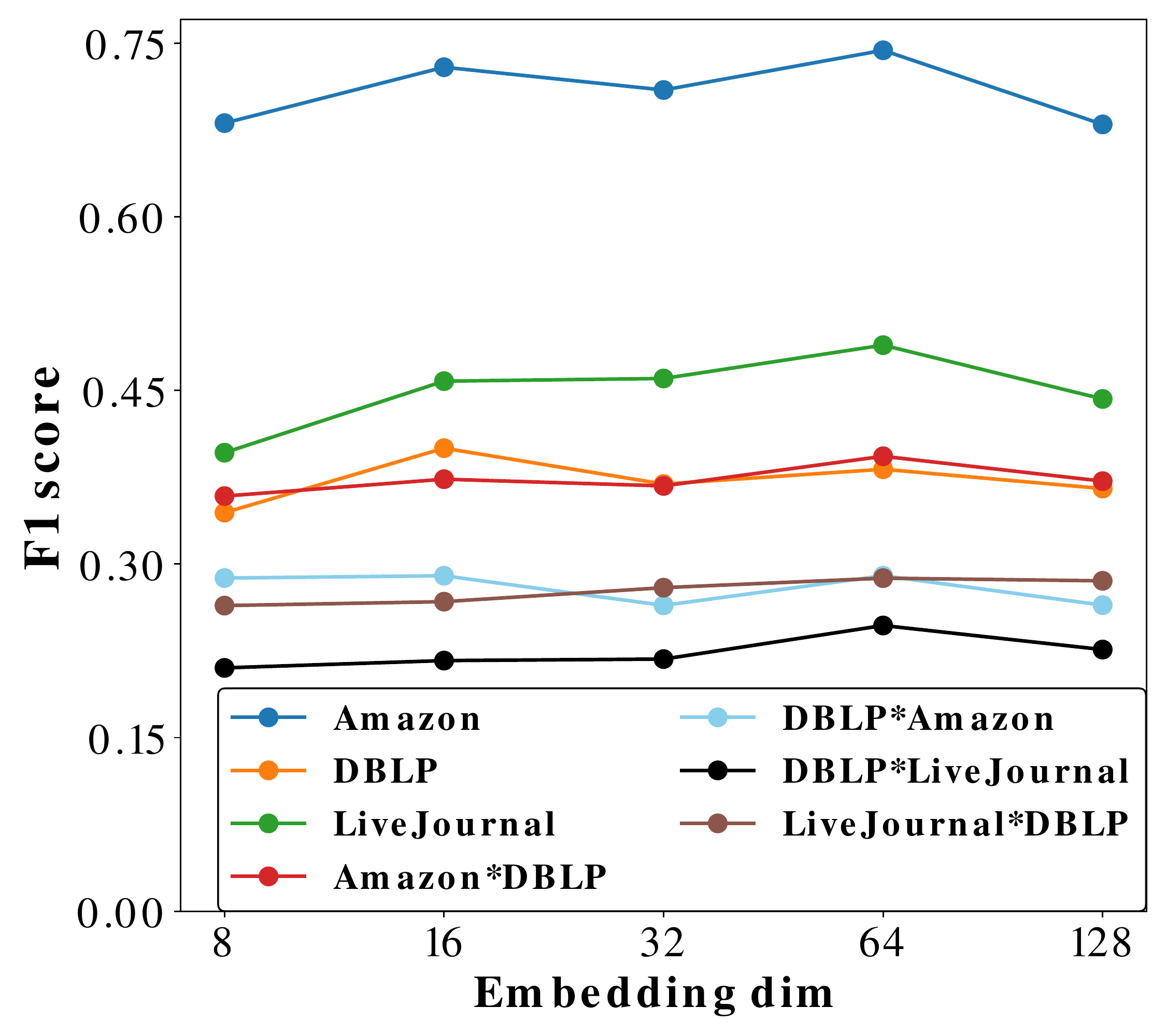}  
      \label{fig:dim}
   \end{minipage}
}
   \subfigure[Different settings of margin $\alpha$]{ 
    \begin{minipage}{4cm}
       \centering   
       \includegraphics[width=4cm]{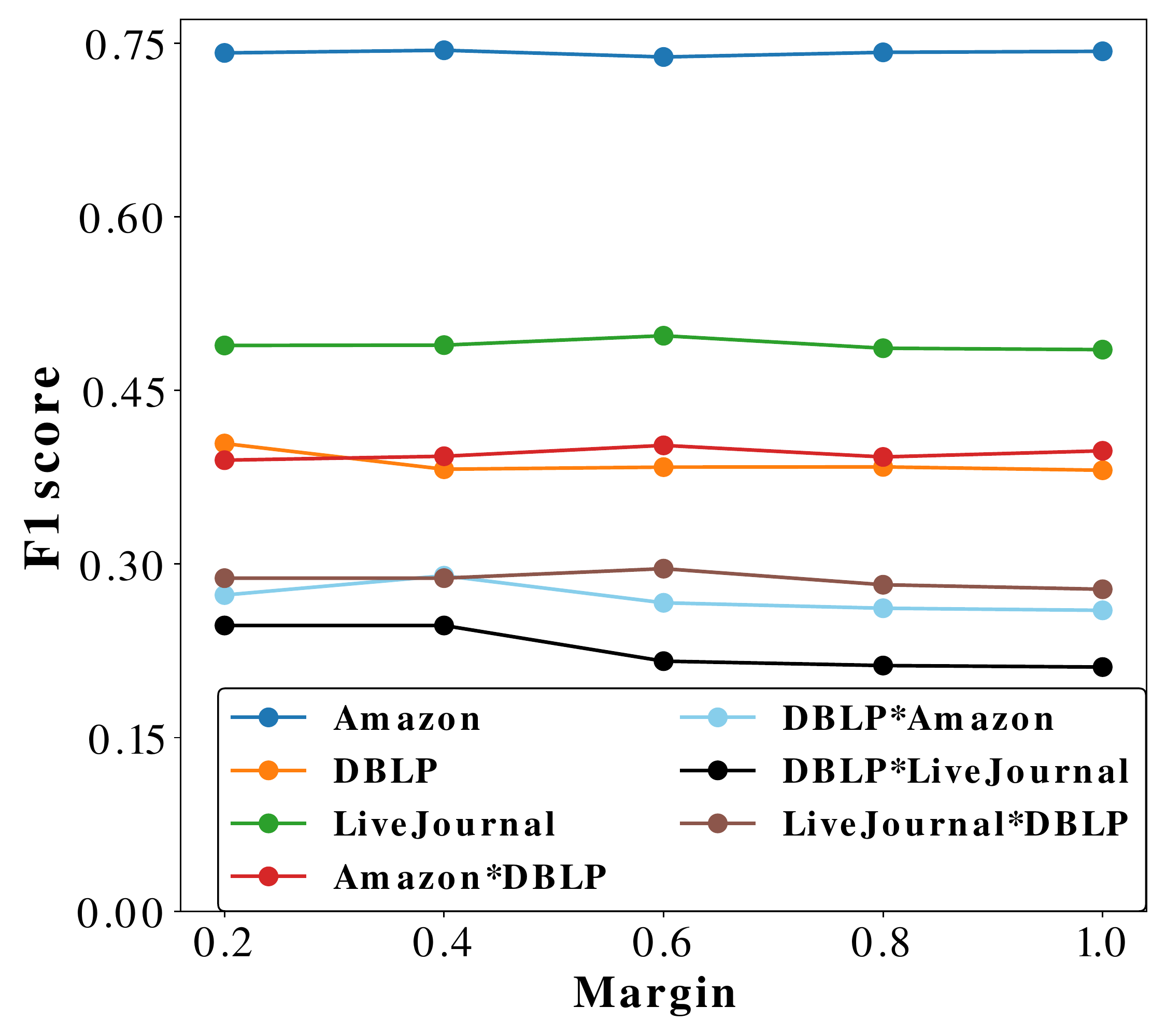}
      \label{fig:margin}
    \end{minipage}
   }
\caption{Parameters study with Locator results reported}   
\label{fig:parameters}   
\end{figure}

 \textbf{Embedding dimension:} Figure \ref{fig:dim} shows the dimension sensitivity experiment results on all datasets. Our model exhibits robustness under different settings of embedding dimensions. In general, \ours \space can reach the peak of F1 score with embedding dimension as 64 on most datasets. Therefore, our model chooses 64 as a standard setting.

\par \textbf{Margin $\alpha$:} As shown in Figure \ref{fig:margin}, our model is quite stable under different settings of $\alpha$ on all datasets. During experiments, we choose 0.4 as a standard setting for all datasets.

\begin{algorithm}
\small
\caption{Community Locator Optimization}
\label{algorithm:1}
\KwInput{A graph $G(\mathcal{V},\mathcal{E},\mathbf{X})$, the set of training communities $\dot{\mathcal{C}}$}
Compute augmented feature matrix $\mathbf{X}'$ based on $\mathbf{X}$

Initialize Encoder $Z$

\While{not converge}
{
   Generate positive pairs $ \mathcal{B}_{pos} = \{ (C^i, C^j)| C^i \subset C^j \subseteq \dot{C}^k \in \dot{\mathcal{C}} \}$
   
   
  Generate negative pairs $ \mathcal{B}_{neg} = \{ (C^i, C^j) |C^i \not\subseteq C^j, C^i \subseteq \dot{C}^p, C^j \subseteq \dot{C}^q, \dot{C}^p, \dot{C}^q \in \dot{\mathcal{C}} \} $
   
  \For{ $\mathrm{each~pair}$ $(C^i, C^j) \in \mathcal{B}_{pos} \cup \mathcal{B}_{neg} $} 
  {
      Encode  $C^i, C^j$ as $\mathbf{z}(C^i)=Z(C^i)$, $\mathbf{z}(C^j)=Z(C^j)$
  }
  
  Traverse over $\mathcal{B}_{pos} \cup \mathcal{B}_{neg}$ to compute $loss$ as Equation\space\ref{eq:distance} 
   
  Update $Z$'s parameters using gradient descent to minimize $loss$
}

\KwOutput{Encoder $Z$}

\end{algorithm}

\begin{algorithm}
\small
\caption{Community Rewriter Optimization}
\label{algorithm:3}
\KwInput{Graph $G(\mathcal{V},\mathcal{E},\mathbf{X})$, encoder $Z$, and training communities $\dot{\mathcal{C}}$ }

Initialize Agent $A$ with parameters $\phi, \psi, \theta$ \\
\While{not converge}
{
  Generate a set of training samples $\mathcal{D}$
 
  \For{ $\mathrm{each~sample}$ $(C^{u} \cup \partial C^{u}, \dot{C}^i)$ $\mathrm{in}$ $\mathcal{D}$}
  {
  Feed forward $C^{u} \cup \partial C^{u}$ to $A$ and obtain the trajectory $\tau$ \\
  Update $\phi, \psi, \theta$ with $\tau$ based on Equation\space\ref{eq:vpg}
  }
}
\KwOutput{Agent $A$}
\end{algorithm}

\begin{algorithm}
\small
\caption{\ours \space Algorithm}
\label{algorithm:smrc}
\KwInput{A graph $G(\mathcal{V},\mathcal{E},\mathbf{X})$, the set of training communities $\dot{\mathcal{C}} (|\dot{\mathcal{C}}|=m)$, and the number of output communities $N$ }

Train Encoder $Z$ according to Algorithm\space\ref{algorithm:1}

$\mathcal{C}^{\text{tmp}}, \hat{\mathcal{C}}  \gets \emptyset, \emptyset$

$n \gets \frac{N}{m}$

Encode training communities as $\dot{\mathbf{Z}} = \{ \mathbf{z}(\dot{C}^i) | \dot{C}^i \in \dot{\mathcal{C}}, i=1,...,m \}$

Encode candidate communities as $ \mathbf{Z} = \{\mathbf{z}(C^u) | u \in \mathcal{V}  \}$

\For{ $\mathrm{each}$ $\mathbf{z}(\dot{C}^i) \in \dot{\mathbf{Z}}$}
{

Find the set of $n$ closest candidate communities in the embedding space $\mathcal{C}$, $\mathcal{C}^{\text{tmp}} = \mathcal{C}^{\text{tmp}} \cup \mathcal{C}$

}

Train Agent $A$ according to Algorithm\space\ref{algorithm:3}

\For{ $\mathrm{each}$ $C \in \mathcal{C}^{tmp}$ }
{
  Feed $C$ to Agent $A$ and obtain a refined community $\hat{C}$

  $\hat{\mathcal{C}} = \hat{\mathcal{C}} \cup \{  \hat{C} \}$

}

\KwOutput{ The set of final predicted communities $\hat{\mathcal{C}}$}

\end{algorithm}

\end{document}